\documentclass[aps,prl,superscriptaddress]{revtex4}
\usepackage{amsmath}
\usepackage{amssymb}
\usepackage{epsfig}
\usepackage{color}
\usepackage{amsmath}
\usepackage[colorlinks]{hyperref}
\usepackage{graphicx,amsmath}
\usepackage[normalem]{ulem}
\usepackage{soul}
\hypersetup{colorlinks,citecolor=red,linkcolor=blue,urlcolor=blue}
\usepackage{graphicx,amsmath}

\begin{document}
\title{Novel Outlook on the Eigenvalue Problem for the Orbital Angular Momentum Operator}
\author{G. S. Japaridze}%\email{gjaparidze@cau.edu}
\affiliation{Clark Atlanta University, Atlanta, GA 30314, USA}\affiliation{Institute of High Energy Physics, Tbilisi State University, Tbilisi 0186, Georgia} \author{A. A. Khelashvili}
\affiliation{Institute of High Energy Physics, Tbilisi State University, Tbilisi 0186, Georgia}
\author{K.~Sh.~Turashvili} \affiliation{Institute of High Energy Physics, Tbilisi State University, Tbilisi 0186, Georgia}

\begin{abstract}
Based on the novel prescription for the power function,  %MDPI: punctuation corrected. here and all over.  
$(x+iy)^m$,  %we present the 
{a} %MDPI: is that the ONLY one, or special?
new expression for $\Psi(x,y|m)$, the eigenfunction of the operator of the third component of the angular 
momentum,
$\hat M_z$, is presented. %MDPI: No "we", "our" recommended for cientific papers unles of privacy or common sense. Here and elsewher below. Please consider the suggestion.
 These functions are normalizable, single valued and, distinct to the traditional 
presentation, 
$(x+iy)^m=\rho^me^{im\phi}$, are invariant under the rotations at $2\pi$ for any, not necessarily integer, $m$ - the eigenvalue of $\hat M_z$. For any real $m$ the functions $\Psi(x,y|m)$ form an orthonormal set, therefore they may serve as a quantum mechanical eigenfunction of  $\hat M_z$. 
 %We report the 
The
eigenfunctions and eigenvalues of the  angular momentum operator squared, 
$\hat M^2$, 
derived for the two different prescriptions for the square root,
 $(m^2)^{1/2}$, $(m^2)^{1/2}=|m|$ and $(m^2)^{1/2}=\pm m$
are reported.
 The normalizable eigenfunctions of  $\hat M^2$ are presented in terms of hypergeometric functions, admitting integer as well as non-integer eigenvalues. It is shown that the purely integer spectrum is not the most general solution but is  just the artifact of a particular choice of the Legendre functions as the  pair of linearly independent solutions of the eigenvalue problem for the $\hat M^2$.
\end{abstract}

\maketitle

\section{1. Introduction}

It is %universally 
{commonly} %MDPI: fits better, please confirm the meaning reatined.
accepted that, from theoretical quantum mechanics, it follows that the spectrum of the eigenvalues of the angular momentum  operator  is discrete and is comprised of the integer values 
only{; %, 
 see,} e.g., \cite{Bohm,Schiff,Fock,LL}. %MDPI: fits better; here and all text. please check.  

Non-integer values of angular momentum do not contradict the principles of quantum theory and were considered few times from different viewpoints. 
Back in {1932,} Majorana 
noted that in the framework of relativistic quantum mechanics, the general formulation of a one-particle equation admits a solution with an arbitrary angular momentum, a predecessor of the theory of infinite dimensional representations of the Lorentz group \cite{M}. 
Working on the analytical properties of the scattering amplitude, {Regge} %MDPI: fits better here. please confirm. here and below.
\cite{Regge} considered angular momentum as a continuous complex 
variable, and derived the singularities in the plane of the complex angular momentum that became universally known as Regge poles.
\mbox{G\"{o}tte et {al}.  \cite{GG}}, %MDPI: ``all'' changed to ``al''. please check all over in the text.
exploiting the freedom in fixing the orientation of phase discontinuity,  introduced states with non-integer angular momentum and 
applied formalism of  the propagation of light modes with the fractional  angular momentum in the paraxial and non-paraxial 
\mbox{regime.}
Exploring polar solutions for the harmonic oscillator,  {Land} \cite{Land} discovered that the Fock space  equivalent to the 
Hilbert space wave 
{functions,} found by solving the Sch\"rodinger equation in spherical coordinates is realized by acting with the creation and 
annihilation {operators,} allowing states with both integer and non-integer  angular momentum. 

In \cite{JKT1,JKT2}, we argued that if only physical conditions are imposed, what can be derived from the principles of quantum mechanics is that the spectrum is discrete with the only condition that the difference $L-|m|$ is integer while $L$ and $m$ could be integer as  well as non-integer. Throughout, $L(L+1)$ is the eigenvalue of the angular momentum operator squared and $m$ is the eigenvalue of the operator of the third component of the \mbox{angular momentum}.

In this {paper,} %MDPI: No "article", "manuscript", "work" etc. recommeded for scientific papers; "study", "paper", "ireview", "investigation" etc to be used instesd. Please consider teh suggestion. here and below. 
 %article, %we report 
a solution of the eigenvalue problem for the  quantum-mechanical orbital angular momentum (hereafter referred to as angular momentum)  operator 
{is reported}  
obtained when only the physical requirement is imposed on the eigenfunction and  is shown that in the framework of theoretical quantum mechanics, the eigenfunctions with both integer and non integer eigenvalues are allowed. 

%This article
 {The paper}  
is organized as follows. In {Section 2}, 
 %we discuss 
the  multivaluedness and periodicity of the eigenfunctions of the operator of the third component of the angular {momentum,} 
$\hat M_z$, {are discussed.}
 % We present a 
{A} 
 new prescription for the power of a complex variable, differing from the  Euler--de Moivre {prescription,} 
$(x+iy)^m=\rho^m\,e^{im\phi}$, used in quantum mechanics, {is presented.}
 Based on this prescription, % we present 
the eigenfunction of $\hat M_z$  in terms of Gauss's hypergeometric series
{is given.} 
 This wave function is normalizable {and} %is %MDPI: no need. extra wording. please confirm.
 distinct from the traditional {eigenfunction being proportional to} 
 %$\sim 
 $e^{im\phi}$, %MDPI: fiths better, otherwise unclear.
%\hl{as well as is shown to be} %MDPI: fits better, please check. 
  is 
single valued and invariant under the rotations at $2\pi$ for 
any, not necessarily integer  $m$. In other words, the requirement of single-valuedness of the wave function does not necessarily lead to the solution with only integer $m$. 
This eigenfunction satisfies the physical requirement of orthonormality, {and, therefore,} 
it can be considered as the wave function describing the physical state with the eigenvalue $m$, 
{being} %MDPI: added, please confirm.
not necessarily integer. 
 In {Section 3}, 
 %we discuss 
{it is discussed} 
how the different prescriptions for the power function alter the eigenfunctions and the spectrum of the  angular momentum operator squared $\hat M^2$. 
 %We present the 
{The} 
eigenfunction of $\hat M^2$ {is found}
which is normalizable and satisfies physical requirements for an integer as well {as} %for a %MDPI: extra wording. please confirm th change. 
non-integer $L$; to a fixed value of $L$  
 {corresponds} %MDPI: very unclear - WHAT is meant here, what is the noun for c"corresponds"? Please rewrite.  
a discrete spectrum of $m$, defined by the relation $|m|=L-k,\,k=\{0,1,\cdots , [L] \}$, where $[L]$ is an integer part of $L$. It is shown that the statement that the spectrum of eigenvalues  consists of only integer $L$ (see, e.g., \cite{Bohm,Schiff,Fock,LL}) is just an artifact of choosing the Legendre 
{function.} 
$P^m_L$, as an eigenfunction of $\hat M^2$. Results are discussed in Section 4.%%%%%%%%%%%%%%%%%%%%%%%%%%%%%%%%%%%%%%%%%%

\section{2. Eigenfunctions of \boldmath{$\hat M_z$} that Are Single-Valued and Periodic for Integer as well as for Non-Integer 
Eigenvalues}
\label{s2}

$\Psi(x,y|m)$, the eigenfunction of  the operator of the third component of the angular {momentum,} $\hat M_z
$,  is defined as the solution of the following eigenvalue equation:
\begin{equation}
\hat M_z \Psi(x,y|m)= i\left(y{d\over dx}-x{d\over dy}\right)\Psi(x,y|m)=m \, \Psi(x,y|m),
\label{eq1}
\end{equation}
where $m$ is the eigenvalue and throughout {the reduced Planck constant} %MDPI: no undefined variables or abbreviations.
$\hbar=1$. Solving Equation~(\ref{eq1}) for the complex  $\Psi(x,y|m)=\Psi_R(x,y|m)+i\Psi_I(x,y|m)$ is  equivalent 
 %of solving 
{to solve} %MDPI: fits better, please confirm the meaning retained.
the following system of the two coupled equations for the real and imaginary parts: 
\begingroup
\makeatletter\def\f@size{9}\check@mathfonts
\def\maketag@@@#1{\hbox{\m@th\normalsize\normalfont#1}}%
\begin{eqnarray}
\label{eq6}
\left( y \frac{d}{ d x} - x \frac{d}{ d y} \right) \Psi_R(x,y|m) = -m \Psi_I (x,y|m) , \quad\quad \left( y \frac{d}{d x} - x \frac{d}{d y} \right) \Psi_I(x,y|m) = m \Psi_R (x,y|m).
\end{eqnarray}
\endgroup  

{Acting} %MDPI: We formatted it as the beginning of the paragraph, please confirm. The same for below.
on 
{Equation} %MDPI: No eq. numbers used solely, "Equation" added. hetre and all over the text. 
(\ref{eq6}) with the operator $(y\,d/dx-x\,d/dy)$ results into the one and the same equation for both $\Psi_R$ and $\Psi_I$:
\begin{eqnarray}
\label{eq7}
\left(x\,{d\over dy}-y\,{d\over dx}\right)^2\Psi_{R,\,I}(x,y|m)=-m^2\,\Psi_{R,\,I}(x,y|m).
\end{eqnarray}

{The} {two linearly independent solutions of the homogeneous differential \mbox{Equation (\ref{eq7})} can be presented as $\Psi_1(x,y|m)=C_1(x; y|m)F_1(x; y|m)$ and $\Psi_2 = C_2(x; y|m) F_2(x; y|m)$, where $F_{1,\,2}$ are linearly independent particular solutions of Equation~(\ref{eq7}) and $C_{1,\,2}$ satisfy the 
{condition,} 
$(yd/dx - x d/dy)C_{1,\,2}  = 0$. 
If %we choose 
{one chooses} 
those $F_{1,\,2}$ that satisfy \mbox{Equation (\ref{eq6})}, then $C_1=C_2$ and the general solution of the eigenvalue Equation (\ref{eq1}) is $\Psi(x,y|m)=C(x; y|m)[F_1(x; y|m)+iF_2(x; y|m)]$, where $C(x; y|m)$ is a complex function 
 %whose  
{with an
absolute value, %is 
fixed}  %MDPI: rewritten; please confirm the meaning retained.
by the physical requirement of normalizability and the phase \mbox{remains undetermined}.}

After %we transform from 
$(x,\,y)$ {is transformed} 
{to} %the %MDPI: extra wording. Eiter "the other" or "another".  
another set of independent {variables,} 
$(f(x^2+y^2)$, $\,\zeta(x,y))$, 
where $f$ is any differentiable function of $x^2+y^2$, Equation~(\ref{eq1}) 
 %will turn 
 {turns} 
into an equation with 
{the only %one 
variable,} %MDPI: fits better. 
$\zeta$.  %We will use this 
{This} %MDPI: No "will" for the current study.
technique of separating variables 
{is used} 
below but first let us quote and discuss the function that is cited in textbooks as a solution {of} %MDPI: No "the" in front of "Eqyuation", "Figure", "Table" when numebreing and not addressing a feature (e.g. "the solution") 
 %the 
Equation~(\ref{eq1}) \cite{Bohm,Schiff,Fock,LL}:
\begin{equation}
F(x,y|m)\sim (x+i y)^m.
\label{eq2}
\end{equation}

%When 
 {If} %MDPI: fits better. please confirm th emeaning retained.
$m$ is %not an 
{non-}integer, %MDPI: replaced, please confirm. 
$F(x; y|m)$ is undetermined, {since} %because 
 for %the %MDPI: fits better.
non-integer {exponents,} the power function is multivalued. 
In order for this function to be a solution of Equation~(\ref{eq1}) it must be defined as a differentiable function of $x$ and $y$. This may be achieved using, e.g., the Euler--de Moivre prescription for the power of a complex number  \cite{ww}:
\begin{eqnarray}
\label{eq3}
(x+i y)^m=(\rho e^{i\phi})^m =\rho^m e^{i m \phi} =\rho^m (\cos \phi+i \sin\phi)^m =\rho^m (\cos m\phi +i \sin m\phi),
\end{eqnarray}
where 
\begin{eqnarray}
\label{eq4}
\rho= |(x^2+y^2)^{1/2}|, \ \sin\phi =\frac{y}{|(x^2+y^2)^{1/2}|},\  \cos\phi =\frac{x}{|(x^2+y^2)^{1/2}|},
\end{eqnarray}
and $|z|$ stands for the absolute value of $z$.

Note that in the chain of Equation (\ref{eq3}) rotational symmetry of the original expression $(x+i y)^m$  is violated when $m$ is 
non-integer. Indeed, due to the invariance of the Cartesian {coordinates,} $x,\,y$, under the 
 %rotation at $2 \pi$, 
 {$2 \pi$-rotation,} % at $2 \pi$, %MDPI: replaced, please confirm the meaning retained. here and below.
$(x+iy)^m$ is formally rotationally invariant for any $m$. Expressions $(\rho e^{i\phi})^m$ and $\rho^m (\cos \phi+i \sin\phi)^m$ are invariant with respect to $\phi\to\phi+2\pi $ for any $m$, while $\rho^m e^{i m \phi}$ and $\rho^m (\cos m\phi +i \sin m\phi)$ violate rotational symmetry for a non-integer $m$. 

If %we require that 
{one requires}
all %the 
 expressions in Equation~(\ref{eq3}) {to} %MDPI: fits better. pleadse confirm the meaning retained.
be invariant with respect to $\phi\to\phi+2\pi $, {then,} $m$ must be an integer. 
{Then,}  $(x+iy)^m$ %becomes 
 {is} %MDPI: fits better, clear enough.
a single valued function 
of $x$ and $y$. This connection between the rotational invariance and the single valuedness of $(x+i y)^m$ 
%\hl{is} %MDPI: missing wording. please confirm.
caused the following assertion: if 
 %we require 
 {one requires} 
the invariance of the wave function $(x+iy)^m\to\rho^me^{im\phi}$ with respect to $\phi\to\phi+2\pi $, this is equivalent to the single valuedness of this function \cite{Bohm,Schiff,Fock,LL}. Both these conditions are satisfied when $m$ is integer and that is 
{a reason} %MDPI: missing word. 
why,  based on the requirements of single valuedness or/and periodicity,  it was declared that $m$ can only be integer and these requirements were formalized in theoretical quantum mechanics as follows \cite{Bohm,Schiff,Fock,LL}:
\begin{eqnarray}
\label{eq5}
\Psi(\rho,\phi|m)=\Psi(\rho,\phi+2\pi k |m),
\end{eqnarray}
where the polar coordinates $\rho$ and $\phi$ are given by 
{Equation} 
(\ref{eq4}) and $k$ is integer.

Imposing these physical conditions of single valuedness and rotational invariance on the wave function that is not observable has been criticized by Pauli  \cite{Pauli} (see also \mbox{in \cite{merz}}). 
We agree with Pauli's criticism and emphasize that the purely integer spectrum of the eigenvalues is obtained only when the conditions of single valuedness and/or periodicity  are realized in the framework of the Euler--de Moivre prescription (\ref{eq3}).  In fact, there are other possible prescriptions for determining $(x+iy)^m$ and it turns out that for one of these 
{prescriptions, the} 
eigenfunction  $\Psi(x,y|m)$ will be single valued, differentiable with respect to its variables, invariant with respect to $\phi\to\phi+2\pi k$ for any $m$, integer as well as non-integer. In other words, even if  
 %we impose
 {one imposes}  
the requirement of single valuedness/rotational invariance on  wave function, this still does not necessarily  lead to a purely integer spectrum. 

To demonstrate {this,} %we use 
 in 
Equation~(\ref{eq7}), 
the technique of separating variables
{is used} 
 %, 
 transforming from $(x,\,y)$ to $(\rho,\,\zeta)$, where $\rho=|(x^2+y^2)^{1/2}|$. 
Now the equation for $F(x,y|m)$  depends only on 
$\zeta$ and %we choose 
$\zeta$ {is chosen} 
so that 
{Equation} 
(\ref{eq7}) %becomes\hl{is}  
 takes a form of an equation solutions of which %{been} %which is  %MDPI: fits better. please confir teh meaning ratained.
are well documented. Defining $\zeta=[1/2 - x/(2|(x^2+y^2)^{1/2}|)  ]=[1/2-x/(2\rho)]$ turns Equation~(\ref{eq7}) into the Gauss 
hypergeometric {equation:} %MDPI: doubling parenthses replaced by the square brakets, please consider the suggestion. here and below.
 \begin{equation}
%\left(
\left[
\zeta(1-\zeta)\,\frac{\rm d^2}{{\rm d} \zeta^2} + \left (  \frac{1}{2}-\zeta \right) \,\frac{\rm d}{{\rm d} \zeta} +m^2  
 %\right) 
 \right] 
F(\zeta|m) = 0 .
\label{eq8}
\end{equation} 

{As} %is well %MDPI: fits better; no "well-known", "well-accepted" etc. to be used in scientific papeers. Please consider the suggestion.  
known, any pair from the Kummer's 24 solutions can be chosen as a set of linearly independent solutions to the 
Gauss equation \cite{ww}; %we choose 
here the {pair,} %:
\begin{eqnarray}
F_1(\zeta|m) &=& {}_2F_1 \left( m,-m;\frac{1}{2}; \zeta\right) = (1-\zeta)^{\frac{1}{2}} \, {}_2F_1 \left( \frac{1}{2}+m,\frac{1}{2}-m; \frac{1}{2}; \zeta\right)  , \nonumber\\
F_2(\zeta|m) &=& \zeta^{\frac{1}{2}} \, {}_2F_1 \left( \frac{1}{2}+m,\frac{1}{2}-m; \frac{3}{2}; \zeta\right) = \zeta^{\frac{1}{2}}  (1-\zeta)^{\frac{1}{2}} \, {}_2F_1 \left( 1+m, 1-m; \frac{3}{2}; \zeta\right) ,
\label{eq9}
\end{eqnarray}
{is chosen},
where ${}_2F_1( a, b; c; \zeta)$ is the Gauss's hypergeometric function \cite{ww,Abramowitz,Br}. Finally, 
$\Psi(x,y|m)=C(F_1+iF_2)$, the eigenfunction of the operator of the third component of the angular momentum, is given 
 {by} %: 
\begin{eqnarray}
\nonumber
\Psi(x,y|m) &=& C(\rho| m) 
 %\left\{ 
 \left[ 
{}_2F_1 \left(m, -m;{1\over 2}; {1\over 2}- {x\over 2 \rho} \right) \right. \\
&&-i m {y\over\rho } \left({1\over 2}+{x\over 2 \rho}\right)^{-1/2} \left. {}_2F_1 \left({1\over 2}+m, {1\over 2}-m;{3\over 2}; {1\over 2}- {x\over 2 \rho}  \right)   
 %\right\},
 \right],
\label{eq10}
\end{eqnarray}
where the square root is determined via prescription $(f^2(x))^{1/2}=|f(x)|$.

 It is straightforward to verify that the eigenfunction (\ref{eq10}),
 {as} a function of $x,\,y$,  is single-valued, invariant under the rotation  at $2\pi k$, is continuous and has continuous 
derivatives of all orders up to infinity for any real, not necessarily integer $m$. Though $\Psi(x,y|m)$ contains a square root, it is infinitely differentiable. This is guaranteed 
 %when  
{as soon as} %MDPI: fits better. please confirm the meaning retained.
$d(x/|x|)/dx=0$ and $d(y/|y|)/dy=0$ are satisfied which is readily demonstrated using 
{Equation} 
(\ref{eq4}). Indeed, from $\cos\phi(x,y)=x/|(x^2+y^2)^{1/2}|$, i.e., $\cos\phi(x,y)|_{y=0}=x/|x|$ it follows {that} %:
\begin{eqnarray}
\label{cos}
&& {d \cos\phi(x,y)\over d x} |_{y=0} = \frac{y^2}{|(x^2+y^2)^{3/2}|} |_{y=0} =0 \quad\to\quad {d(x/|x|)\over dx}=0.
\end{eqnarray}

{Similar} to 
{that of} %MDPI added, otherwise unclear. please confirm the meaning reatined.
{Equation} (\ref{cos}), from $d\sin\phi(x,y)/dy|_{x=0}$: 
 %we obtain 
$d(y/|y|)/dy=0$. %MDPI: extra wording.

{Let us consider particular values of $m$. %We 
 {Let us} 
start %from
 {with} %MDPI: fits better. 
the integer}  $m=\pm N$, \mbox{$\,N=1,  2,  3\cdots$}. 
{{The corresponding} %Corresponding 
hypergeometric functions} ${}_2F_1(N,-N;1/2;z)$ and ${}_2F_1(1/2+N,1/2-N;3/2;z)$ are tabulated (see, e.g., \cite{Abramowitz}){.
 % and  
 Then,} 
 %from 
Equation~(\ref{eq10}) {reads:} %it follows: %MDPI: rewritten, broken. please confirm.
\begin{eqnarray}
\Psi(x,y|\pm N)={C(\rho|\pm N)\over \rho^N}\,(x\mp i y)^N.
\label{eq11}
\end{eqnarray}
 
 {So}, for {integer} $m$, $\Psi(x,y|N)$ reproduces, up to the factor $C(\rho|\pm N)/\rho^N$,  solution (\ref{eq2}),  
\mbox{$(x+i 
y)^N$}.

%Next we consider 
{Let us consider next} 
the half-integer values of $m$. Explicit expressions are lengthy and involved; 
 %we will give 
results for $m=\pm 1/2$ and $m=\pm 3/2$ {are:}
\begin{eqnarray}
\Psi\left(x,y|\pm \frac{1}{2}\right) &=&C\left(\rho|\pm \frac{1}{2}\right)  
 %\left\{ 
 \left[ 
\left( \frac{1}{2} +\frac{x}{2\rho}\right)^{\frac{1}{2}}  \mp i \frac{y}{2\rho}   \left( \frac{1}{2} +\frac{x}{2\rho}\right)^{ -\frac{1}{2}} 
 %\right\}  ,
 \right]  ,
 %\nonumber\\
\label{eq16}\\
\Psi\left(x,y|\pm \frac{3}{2}\right)&=&C\left(\rho|\pm \frac{3}{2}\right)  
 %\left\{ 
\left[ 
\left( \frac{1}{2} +\frac{x}{2\rho}\right)^{\frac{1}{2}} \left( \frac{2x}{\rho} -1\right)  \mp i  \frac{y}{2\rho}   \left( \frac{1}{2} +\frac{x}{2\rho}\right)^{ -\frac{1}{2}}  \left( 1 +\frac{2x}{\rho}\right) 
 % \right\}\nonumber.
 \right]\nonumber.
\end{eqnarray}

 %We found a 
%\hl{Above, a} 
A relation between the wave functions for the integer and half-integer $m$  
{is found being}  %which we have 
verified for $m=1/2, 3/2,5/2$:
\begin{eqnarray}
\left [{\Psi\left(x,y|\pm \frac{N}{2}\right)\over C\left(\rho|\pm \frac{N}{2}\right)} \right]^2  = { \Psi(x,y|\pm N)\over  C\left(\rho|\pm N\right) },   
\label{eq18}
\end{eqnarray}
i.e., wave function for the half-integer $m$, $\Psi(x,y|\pm N/2)$, satisfies $\Psi^2(x,y|\pm N/2)\sim \Psi(x,y|\pm N)$, a relation similar to that which holds for 
{Equation} 
(\ref{eq2}), $((x\pm i y)^{N/2})^{2}=(x\pm i y)^{N}$. This result, 
 %together 
{along} %MDPI: fits better.
with 
{Equation} 
(\ref{eq11}), indicates that the eigenfunction $\Psi(x,y|m)$, given by Equation~(\ref{eq10}), presents one possible prescription for the power function $(x+iy)^m$. 

Let us demonstrate with the example of the half-integer $m$ the importance 
 of 
%\hl{for} %MDPI: replaced, please confirm.
choosing prescription for  the square root as $(f^2(x))^{1/2}=|f(x)|$. 
 %To do so we change 
{To this end, one moves}   %MDPI: fits better.
from Cartesian to polar coordinates, $(x,y)\to (\rho, \phi)$,  see Equation~(\ref{eq4}). In polar
 {coordinates,  
the argument,} 
$(1/2-x/2\rho)$,  %MDPI: moved here, fits better; please confirm the meaning retained.
of the hypergeometric 
functions 
 %becomes 
{reads:} 
$(1-\cos\phi)/2 = \sin^2(\phi/2)$.  
{First,} %we use 
the {prescription,}  $(f^2(x))^{1/2}=f(x)$ {is used.}
 Using in 
{Equation} 
(\ref{eq16}) the %well 
{known} relation   ${}_2F_1(a,b; 
b;z)=(1-z)^{-a}$ \cite{Abramowitz,Br}, {one obtains:}
 %we obtain:
\begingroup
\makeatletter\def\f@size{9}\check@mathfonts
\def\maketag@@@#1{\hbox{\m@th\normalsize\normalfont#1}}%
\begin{eqnarray}
\Psi\left(x,y |\pm \frac{1}{2}\right)&\to&\Psi\left(\rho,\phi |\pm \frac{1}{2}\right)=  C\left(\rho|\pm \frac{1}{2}\right)  
%\left\{ 
\left[ 
\left( 
\cos^2\frac{\phi}{2}
%\right]^{\frac{1}{2}}  
\right)^{\frac{1}{2}}  
\mp \frac{i}{2}  \,\sin\phi  
 %\left[ 
 \left( 
\cos^2\frac{\phi}{2}
 %\right]^{-\frac{1}{2}}  \right\}=\cr\cr
 \right)^{-\frac{1}{2}}  
   %\right\}
   \right]
=\cr\cr
&&C\left(\rho|\pm \frac{1}{2}\right)  
 %\left\{  
\left(  
\cos\frac{\phi}{2}   \mp i  \,\sin\frac{\phi}{2}   
 %\right\} 
 \right) 
=  C\left(\rho|\pm \frac{1}{2}\right)   \,e^{\mp i \frac{\phi}{2}  }.
\label{eq19}
\end{eqnarray}
\endgroup

{Equation}~(\ref{eq19}), apart from the normalization factor, is the 
 %well 
 {known}  
$e^{im\phi}$, $m=\pm 1/2$, originated by the Euler--de Moivre prescription for the $(x+iy)^{1/2}$ and presented as a standard expression for the eigenfunction of $\hat M_z$  \cite{Bohm,Schiff,Fock,LL}. 
 %Clearly,  %MDPI: No "clearly", "undobtedly" etc. recommende dfor scientific papers. Please consider the suggestion.
 {Obviously,} 
because of the rotational invariance of Cartesian coordinates, $x(\phi)=x(\phi+2k\pi),\,y(\phi)=y(\phi+2k\pi)$, the left hand side of  Equation~(\ref{eq19}),  $\Psi(x,y|\pm 1/2)$, given by Equation~(\ref{eq16}), is invariant under the rotation  $\phi\to \phi + 2 \pi k$. On the other hand, the 
{right-hand side (r.h.s.)} %MDPI: the abbreviation introduced for further use; the dash added. 
of 
{Equation} 
(\ref{eq19}) is invariant under the  translations $\phi\to \phi + 4 \pi k$, %and 
 {but} %MDPI: please confirm the replacement.
not under $\phi\to \phi + 2 \pi k$. This inconsistency stems from %using 
{the %MDPI: fits better. 
prescription,} $(f^2(x))^{1/2}=f(x)$ while deriving Equation~(\ref{eq19}). For example,  $\cos(\phi/2)$ appeared in the real part  
of 
{Equation} 
(\ref{eq19}) 
because for $(\cos^2(z))^{1/2}$ we used $\cos(z)$:
 %{as soon as is used} 
%$\cos(z)$ for $(\cos^2(z))^{1/2}$:
\begin{equation}
{}_2F_1\left(
-{1\over 2},{1\over 2};{1\over 2}; \sin^2\left({\phi\over 2}\right)
\right)
=
 %\left(
 \left[
1-\sin^2\left({\phi\over 2}\right)
 %\right)^{1/2} 
 \right]^{1/2} 
= 
 %\left(
 \left[
\cos^2\left({\phi\over 2}\right)
 %\right)^{1/2}
 \right]^{1/2}
=\cos\left({\phi\over 2}\right).
\label{eq21}
\end{equation}

%When 
 {If} 
$\phi =2 \pi k$, the 
{left-hand side (l.h.s.)}  
of this relation is unity,  ${}_2F_1(-1/2,1/2;1/2;0)=+1$, while for 
 %the right hand side 
 {r.h.s.} 
 %we have 
{one gets}
$\cos \pi k$, which, depending on $k$, can be either $+1$ or $-1$.
 % When 
 {If} 
 $\phi=\pi (2 k+1) $, both 
 %left and the right hand sides 
 {l.h.s. and r.h.s.} 
of {Equation} 
(\ref{eq21}) vanish. This means that Equation~(\ref{eq21}) is 
valid only for %those 
$\phi$ 
 %for which
 {with} %MDPI: rewritten. please confirm.  
$\cos(\phi/2)\geq 0${;  %(
this} condition, lacking from \cite{Abramowitz,Br}, is also noted in Ref.~\cite{NIST}. %).
 %At the same time, 
 {Meantime,}
both 
 %left and right hand sides 
 {l.h.s. and r.h.s.} 
of Equation~(\ref{eq21}) exist and are well defined for all values of $\phi$ and that calls for the question of how  relation (\ref{eq21})  has to be interpreted when $\cos(\phi/2)<0$. Note that the inconsistency does not 
{arise}  
 %at all  %MDPI: extra wording.
if, as an alternative, 
instead of {Equation} (\ref{eq21}), 
 %we use the 
{the} 
following {relation,} %:
\begin{equation}
{}_2F_1\left(-{1\over 2},{1\over 2};{1\over 2}; \sin^2\left({\phi\over 2}\right)\right)=\left|\cos\left({\phi\over 2}\right)\right|,
\label{eq22}
\end{equation}
{is used,} 
i.e., if %we use
 {one applies the}  
prescription $(\cos^2(\phi/2))^{1/2}=|\cos(\phi/2)|$, %and 
 {but} 
not $(\cos^2(\phi/2))^{1/2}=\cos(\phi/2)$. 

 %When 
{If,}
instead of the {prescription,} $(f^2(x))^{1/2}=f(x)$,  %we use 
$(f^2(x))^{1/2}=|f(x)|$ {is applied, one obtains:} %, \mbox{we obtain}:
\begin{eqnarray}
\Psi\left(x,y |\pm \frac{1}{2}\right)\to\Psi\left(\rho,\phi |\pm \frac{1}{2}\right)=C\left(\rho|\pm \frac{1}{2}\right)  
 %\left\{ 
\left( 
\left|\cos\frac{\phi}{2}\right|   \mp {i\over 2}\, \sin\phi  \left|\cos\frac{\phi}{2}\right|^{-1}   
 %\right\}\ . 
 \right)\ . 
\label{eq191}
\end{eqnarray}

{Expression} (\ref{eq191}) is well defined  for all values of $\phi$ and, most importantly, it is invariant under 
{translations,} 
$\phi\to \phi + 2 \pi k$; the 
{above-mentioned} %MDPI: dash added.
inconsistency disappears.   Similarly, 
using, 
for the case {of} %MDPI: lost wording.
$m=3/2$, the {prescription,} $(f^2(x))^{1/2}=|f(x)|$, 
 %we obtain 
{one obtains} 
the same result, $\Psi(\rho,\phi | \pm 3/2 
)=\Psi(\rho,\phi+2k\pi | \pm 
3/2 )$ %while 
{but} %MDPI: seems better, please confirm the meaning retained.
for the {prescription,} $(f^2(x))^{1/2}=f(x)$, the resulting wave function is no longer invariant under $\phi\to \phi+2\pi$.

From  Equation~(\ref{eq1}) and its conjugation, using the properties of the Gauss hypergeometric functions  (see, e.g., \cite{ww}),  
 %we derived 
{one finds} 
that $|\Psi(x,y|m)|^2$ %depends only
 {only depends} %MDPI: fits better. 
on \mbox{$\rho=|(x^2+y^2)^{1/2}|$} {and} %that  %MDPI: extra wording.
the real and imaginary parts of $\Psi(x,y|m)$ satisfy the relation resembling the trigonometric {identity,} $\cos^2 x+\sin^2 x 
= 1$:
\begin{eqnarray}
{\Psi^2_R(x,y|m)+\Psi^2_I(x,y|m)\over |C(\rho|m)|^2}=1. %,
\label{eq25}
\end{eqnarray}
%where %$\Psi_{R,\,I}$  are the real and imaginary parts of $\Psi$. %MDPI: No need, already defined for Eq.(3). 

{Particular} examples of the general result 
(\ref{eq25}) are cases 
of integer $m=N$, when {Equation} 
(\ref{eq25})  reduces to $\cos^2 N\phi+\sin^2 N\phi = 1$, and of half-integer $m$, 
 %we quote 
 {quoted} here 
 %case  
{for} $m=1/2$: %$,  %MDPI: rewrittem=n fits better. please onfirm the emaning retained.
$(\Psi^2_R(x,y|1/2) + \Psi^2_I (x,y| 1/2) )/ |C(\rho| 1/2)|^2= 
|\cos(\phi/2)|^2+\sin^2(\phi)/(4|\cos(\phi/2)|^{2})=1$. Relation (\ref{eq25}) is another indication that the functions (\ref{eq2}) 
and (\ref{eq10}) belong to the same {class} %, 
since for any $m$, %they 
{both Equations} (\ref{eq2}) and (\ref{eq10})   %MDPI: please confirm.
satisfy relation $|\Psi/C|^2=1$.
 
The physical requirement the solution should satisfy is that  $\Psi(x,y|m)$ must be orthonormal.
To verify {normalizability,  let us} we 
use relation (\ref{eq25}). Normalizability {condition,}
\begin{eqnarray}
\int_{-\infty}^{\infty}\int_{-\infty}^{\infty}{\rm d} x \,  {\rm d} y \, |\Psi(x,y|m)|^2 = \pi \int_{0}^{\infty}{\rm d}  \rho\, |C(\rho|m)|^2 < \infty
\label{eq26}
\end{eqnarray}
 can be readily realized by the appropriate choice of $C$.  It  suffices to choose $|C|^2\sim \rho^\gamma$ with $\gamma< -1$ 
 %when 
{for} 
$\rho\to\infty$ and $C$ finite 
{for} 
 %when  %MDPI: please confirm th emeaning retained.
$\rho\to 0$.

Orthogonality follows from the relation which is obtained from 
 %the 
{Equation}~(\ref{eq1}):
 \begin{eqnarray}
&&  i(m'-m)\int\int_{-\infty}^{\infty}{\rm d}  x\,  {\rm d} y \, \Psi^*(x,y|m') \Psi(x,y|m) \nonumber\\
&& \quad = \int_{-\infty}^{\infty}  {\rm } {\rm d} y \, y \Psi^* (x,y|m') [ \Psi (x,y|m)|_{x=\infty} - \Psi (x,y|m)|_{x=- \infty} ] \nonumber\\
&&\quad - \int_{-\infty}^{\infty}  {\rm } {\rm d} x \, x \Psi^* (x,y|m') [\Psi (x,y|m)|_{y=\infty} - \Psi (x,y|m)|_{y=- \infty}].
\label{eq28}
\end{eqnarray} 

{Using} {the {above-mentioned}  constraints on  $C(\rho|m)$ 
%we obtain} 
{one obtains}
$\Psi (x,y|m)|_{x,y=\infty} - \Psi (x,y|m)|_{x,y=- \infty}  =0$, from which %it 
 {follows} that, if the integral in 
 %the left hand side 
{l.h.s.} 
of Equation~(\ref{eq28}) exists, {then,} 
for $(m' - m)\neq 0$, %it
 {the integral} %MDPI: rewritten; unclear otherwise. Please confirm teh meaning retained. 
is zero {being} %, which is 
the condition of orthogonality. 
Therefore, $\Psi (x,y|m)$ from 
{Equation} 
(\ref{eq10})  fulfills every physical requirement  
{which} %MDPI: pleae confirm 
the eigenfunction of the third component of the quantum mechanical angular momentum operator should satisfy. 

Using a certain prescription  for  the power function may lead to an expression of the wave function that is not invariant under the translations at $2k\pi$ for a non-integer eigenvalues. This case is  realized by the Euler--de Moivre 
{prescription,} $(\rho e^{i\phi})^m=\rho^m e^{im\phi}$. On the other hand, if 
 %we use 
another prescription {is applied,}
 this may result in a wave function that is invariant under the translations at $2k\pi$ for integer as well as for non-integer 
eigenvalues. This case is realized by the eigenfunction (\ref{eq10}), 
{where the prescription,} $(f^2(x))^{1/2}=|f(x)|$, is used. 
Prescription for the power function {affects 
not only the features 
of eigenfunctions of} $\hat M_z$, but also  {the features of} %MDPII: rewritten, fits better. please confirm.
the eigenvalues of the operator of the angular momentum squared, {as} %is %MDPI: extra wording. 
described in 
%the following section.
{Section}~\ref{s3} below. %MDPI: always recommended to number Sections.
%%%%%%%%%%%%%%%%%%%%%%%%%%%%%%%%%%%%%%%%%%
\section{3. Eigenfunctions and Eigenvalues of \boldmath{$ \hat M^2$}}
\label{s3}

The eigenvalue equation  for $\hat M^2=\hat M_x^2+\hat M_y^2+\hat M_z^2$, the operator of the angular momentum {squared,} %:
\begin{equation}
\label{M2}
\hspace{-20mm} \hat M^2\Psi_{M}(L,m|\theta)=\left(\sin^2\theta\,{d^2 \over d \cos^2\theta}-2 \cos\theta { d \over d \cos\theta} - \frac{m^2}{\sin^2\theta}\right)\Psi_{M}(L,m|\theta) = L(L+1)\Psi_{M}(L,m|\theta),             %\\ \nonumber
\end{equation}
where $\theta$ is {the} polar 
angle and $L>0$ is the eigenvalue, 
 %also 
 {reduces} %MDPI: why "also"? Removed, please check. 
to the equation for the Gauss's hypergeometric series, solutions of which can be represented by various linearly independent  pairs of functions. 
 %One 
 {A}  %MDPI: replaced. please confirm.
possible pair is $\Psi_{M(1)},\;\Psi_{M(2)}$ \cite{JKT1}: 
 %\begin{eqnarray}
 \begin{eqnarray}
  \Psi_{M(1)}(L,m|\theta) &=& f_1(m|\theta) \,\Phi_1(L,m|\theta ), \quad  f_1(m|\theta) =(\sin^2\theta) ^ {{(m^2)^{1/2}\over 2}}, \nonumber\\ 
 && \Phi_1(L,m|\theta ) ={}_2F_1\left({1\over 2}+{(m^2)^{1/2}\over 2}+{L\over 2},{(m^2)^{1/2}\over 2}-{L\over 2};{1\over 2};\cos^2\theta\right) ; \nonumber\\[0.2in]
 \Psi_{M(2)}(L, m|\theta) &=& f_2(m|\theta) \,\Phi_2(L,m|\theta ), \quad  f_2(m|\theta) = \cos\theta\,(\sin^2\theta) ^ {{(m^2)^{1/2}\over 2}}, \label{eq30}\\
 && \Phi_2(L,m|\theta ) ={}_2F_1\left(1+{(m^2)^{1/2}\over 2}+{L\over 2}, {1\over 2}+{(m^2)^{1/2}\over 2}-{L\over 2};{3\over 2};\cos^2\theta\right).\nonumber
%\label{eq30}
\end{eqnarray}
Any linear superposition of $\Psi_{M(1)}(L, m|\theta)$ and $\Psi_{M(2)}(L, m|\theta)$ is also a solution 
 {of} 
 %the 
Equation~(\ref{M2}).

The only physical requirement for the functions $\Psi_{M(1,\,2)}(L,m|\theta)$ is 
{the} normalizability. The necessary condition for the normalizability is that $\Psi_M(L,m|\theta)$, presented as 
a {product,} $\Psi=f\Phi$, must 
be a 
regular function.  Unfortunately it is not known how to realize such a condition for the product and the only option left is that normalizability can be achieved when the factors  
$f_j(m|\theta)$ and $\Phi_j(L,m|\theta )$ are regular for {all}  
 %the  
values of their arguments which 
 %will lead 
 {leads} 
to a regular product $\Psi=f\Phi$. 

{Functions,}  $f_j(m|\theta)\sim(\sin^2\theta) ^{(m^2)^{1/2}/2}$, are regular for any $\theta$ when $(m^2)^{1/2}\geq 0$. Hypergeometric 
{series,} 
${}_2F_1(a,b; c;\cos^2\theta)$, converge for $\cos^2\theta<1$; %when
 {for}  
$\cos^2\theta=1$ %they 
 {the series} %MDPI: please confirm the memaning.
converge only if  $a+b-c<0$ \cite{ww}. This condition 
for ${}_2F_1$  from Equation~(\ref{eq30}) {reads:} $a+b-c=(m^2)^{1/2}<0$, which is opposite to the condition of {the regularity of} % of 
$f_j(m|\theta)$,  
$(m^2)^{1/2}\geq 0$.
Therefore, $f_j(m|\theta)$ and infinite series ${}_2F_1$ cannot 
{simultaneously 
be
regular, and }  %and The sentece rewritten, breaks into two. please confirm the emaning retained. 
in order} %for 
$\Psi_{M(j)}(L,m|\theta)$ to be regular, the infinite hypergeometric series ${}_2F_1$ should {terminate} %, 
resulting in 
polynomials; {hereafter,}
this truncation of infinite hypergeometric series 
 %hereafter 
 is referred to as {``polynomialization.''}
  As known, the infinite hypergeometric series ${}_2F_1(a;b; c;z)$  %become 
 {are} %MDPI: fits better.
polynomials %when 
{if} 
either 
$a$ or $b$ is a non-positive integer \cite{ww}. Since the parameters of ${}_2F_1(a,\,b;\,c;\;\cos^2\theta)$ depend on $L,\,m$ (see Equation~(\ref{eq30})), setting $a$ or $b$ to a non-positive integer results into constraints on $L,\,m$. 
 %Clearly, 
  {It is obvious} that a different  prescription for $(m^2)^{1/2}$ generates different restrictions on eigenvalues of the angular momentum.

Let us report results for the eigenvalue problem of the operator of the angular momentum squared{; %(
 for  %the 
 explicit but somewhat} %MDPI: fits better. 
 lengthy {calculations,} see \cite{JKT1}. %). 
 It is  essential to specify which prescription is used for the $(m^2)^{1/2}$ appearing in $\Psi_{M(j)}(L,m|\theta)$, given by Equation (\ref{eq30}), since, as shown below,  after the normalizability is required, different prescriptions lead to the different results for the spectrum. 
 %{To stress is that it} %It 
% is %important 
 % {essential} to specify %which
% {the}  prescription %is
 %used {for} %the 
 %$(m^2)^{1/2}$, appearing in 
% $\Psi_{M(j)}(L,m|\theta)$, given by {Equation} (\ref{eq30}), {since, as} %it is 
% shown below,  after 
 %requiring 
%{the} normalizability {is required,} 
 %the 
 %different prescriptions lead to %the 
% different results for the spectrum. %MDPI: rewritten, fits better.

 There are two possible {prescriptions, namely,} %:
 $(m^2)^{1/2}=|m|$ and $(m^2)^{1/2}=\pm m$. 
 First, %we consider 
the case when the prescription is $(m^2)^{1/2}=|m|$ {is considered.} Equating parameters %$a,\,b$  
 {$a$ and $b$} %MDPI: when two varaibales listed, the "and" fits better.
of the two hypergeometric 
functions from Equation~(\ref{eq30}) to  non-positive integers, $-k$, results in four conditions, two 
{conditions} 
for $\Psi_{M(1)}$ and two {conditions} for $\Psi_{M(2)}$. One condition out of {the} 
two for  $\Psi_{M(1)}$ generates singular functions and, {thus,} 
 %has 
to be 
dropped; the same is true for $\Psi_{M(2)}$ {and, finally, one remains with} only two conditions of polynomialization, generating regular 
{eigenfunctions}~\cite{JKT1}.  %MDPI: fits better. please confirm. 
 %, are left 
 
The spectrum of eigenvalues corresponding to {the %se 
two left} regular eigenfunctions is obtained from the following polynomialization conditions:
 \begin{eqnarray}
 \label{eq31}
&&\left|{m_{(1)}\over 2}\right|-{L\over 2}= - k_1;  \ L - 2\left[{L\over 2}\right] \leq |m_{(1)}| \leq L, 
\end{eqnarray}
\begin{eqnarray}
\label{eq310}
&&\left|{m_{(2)}\over 2}\right|-{L-1\over 2}= - k_2;  \ (L-1) - 2\left[{L-1\over 2}\right] \leq |m_{(2)}| \leq (L-1).
\end{eqnarray}  

{Here,} $[X]$ stands for the integer part of $X$ satisfying $X-[X]\geq 0$, $k_{1}=0,1,2,\cdots[L/2]$ and $k_{2}=0,1,2,\cdots[(L-1)/2]$. The 
sets (\ref{eq31}) and (\ref{eq310}) are comprised of the numeric sequence of positive and negative elements $m_{(j),k},\;j=1,2$ with the step 
size 2, e.g., $|m_{(1)}|=L-2k_1$. From Equations~(\ref{eq31}) and (\ref{eq310}) it follows that the spectrum is discrete with the only condition 
that $L-|m|$ is necessarily {integer,} 
while there are no constraints  on $L$ and $m$ separately; the solution (\ref{eq30}) is regular for integer as  well as for non-integer $L,\,m$.

The sets of {the 
eigenvalues,} $\{m_{(1)}\}$ and $\{m_{(2)}\}$ (and their corresponding {eigenfunctions),} can be formally %united 
{combined} %MDPI: fits bettr.
into one {set,} comprised of the 
positive and negative {elements,} $m_k$, with the step size {1:} %, 
 $m_{k}-m_{k-1}=\pm 1$. Using numerical ordering from the smallest to the 
largest value, {the combined} %is united 
set of all possible eigenvalues %is presented 
{reads} as follows:
\begin{eqnarray}
\{m\}|_{(m^2)^{1/2}=|m|} & =&
\{-L,-L+1,-L+2,.., -m_0 ; m_0,..,L-2,L-1,L \},
\label{eq1.20}
\end{eqnarray}
where, depending on  a numeric value of $L$, $m_0$ is either $(L-2[L/2])$, the minimal positive value  from the set (\ref{eq31}), or $(L-1-2[(L-1)/2])$, the minimal positive value from the \mbox{set (\ref{eq310}) \cite{JKT1}}. 

Starting from the subset  of 
{Equation} 
(\ref{eq1.20}) with $m$ positive, applying $\hat M_{-}=\hat M_{x}-i\hat M_{y}$ leads to a subset with the negative $m$ and vice versa, $\hat 
M_{+}\Psi(m<0)=(\hat M_{x}+i\hat M_{y})\Psi(m<0)\rightarrow \Psi(m>0)$ only when $L$ 
{is}  %MDPI: the arrow symbol  (LaTeX $to$) is NOT defined and is unclear. Please define it or use wording instead.
either integer or half-integer. Acting by $\hat M_{-}$ on the 
regular functions with $m$ positive leads to the regular functions with $m$ negative, $\hat M_{-}\Psi_{\mathrm{reg}}(m>0)\to 
\Psi_{\mathrm{reg}}(m<0)$ only when $L$ is integer. When $L$ is half-integer, acting by $\hat M_{-}$ on the regular functions with $m$ positive 
leads to the singular functions with $m$ negative, $\hat M_{-}\Psi_{\mathrm{reg}}(m>0)\to \Psi_{\mathrm{sing}}(m<0)$. A symmetric result is valid 
when applying the rising operator: $\hat M_{+}\Psi_{\mathrm{reg}}(m<0)\to \Psi_{\mathrm{reg}}(m>0)$ only when $L$ is integer and when $L$ is 
half-{integer,} $\hat M_{+}\Psi_{\mathrm{reg}}(m<0)\to \Psi_{\mathrm{sing}}(m>0)$ \cite{JKT1}.

 %Clearly, 
 {Evidently,} 
if it is required that when %we require that  
 %{the} 
moving  with the step size 1,
 %{is required,} 
 starting from the wave function with $m=(-L)$, one  should arrive at the wave function with $m=+L$ and vice 
versa, this will be possible only when $m$ is either integer or half-integer. In this case, no analysis of the eigenvalue problem is necessary since the spectrum is already predefined to consist of only integer or half-integer $m$. 
The requirement {that,} starting from the state with $m=\mp L$ %meaning 
 %we arrive, 
{one arrives,} 
moving  with step size 1, to the state with $m=\pm L$, is postulated in the method of commutator algebra of the angular momentum operators \cite{Bohm,Schiff,Fock,LL}. 

{This requirement,} %MDPI: The sentence is too long and far to be clear. Please rewrite suggesting to be separated into two sentences. 
customarily taken for granted to be a physical postulate, is actually a mathematical {condition,} 
imposed by hand which filters out possible {non-integer} and {non-half-integer} $m$ from the spectrum, similarly to how imposing the non physical 
condition 
of 
periodicity on $e^{im\phi}$, {filters} out non integer $m$ from the spectrum of $\hat M_z$.  When $L$ is non-integer, the requirement that starting from 
the state with $m=\mp L$ 
 %we arrive 
{one arrives} 
at the state with $m=\pm L$, cannot be satisfied. Indeed, {e.g.,} %say %MDPI: jargon. 
for $L=1.7$, acting with the lowering operator $\hat M_-=\hat M_x-i\hat M_y$ on $\Psi(1.7,1.7|\theta)$ would never result in  $\Psi(1.7,-1.7|\theta)$ and then 
{terminate;}  %stop; please confirm teh meaning retained. 
 instead 
 %we now have 
{one gets:} 
$\Psi(1.7,1.7|\theta)\rightarrow \Psi(1.7,0.7|\theta)\rightarrow \Psi(1.7,-0.3|\theta)\rightarrow \Psi(1.7,-1.3|\theta)\rightarrow \Psi(1.7,-2.3|\theta)\rightarrow\cdots $. Let us recall that as an alternative to an unphysical requirement of  single valuedness  of the wave function, Pauli suggested that acting by the rising and lowering operators $\hat M_x\pm i \hat M_y$ on regular wave functions 
 %we should have 
{one should find:}
$\Psi_{M}(L,\,-L|\theta)\leftrightarrow\Psi_{M}(L,\,-L+1|\theta) \leftrightarrow\cdots\leftrightarrow\Psi_{M}(L,\,-1+L|\theta)\leftrightarrow \Psi_{M}(L,\,L|\theta)$. 
 %He
 {Pauli} %MDPI: No "he", "she", "it", "they" recommened in scientific papers, especially addressing persons. 
justified this by postulating that as a result of acting on  regular wave functions by $\hat M_x\pm i \hat M_y$, no singular functions 
 %should 
\mbox{{appear} \cite{Pauli,merz}}. %MDPI: fits better.
In the case of the prescription $(m^2)^{1/2}=|m|$,  moving up and down in spectrum with %steps
 {step} %MDPI: singular. 
size 2, 
indeed no singular functions are generated for any, integer or {non-integer,} $m$, 
{as} follows from the conditions of  polynomialization %MDPI: fits better here.  
 \cite{JKT1}. 

However,  singular functions appear if instead {of} 
\mbox{$(m^2)^{1/2}=|m|$} the  prescription $(m^2)^{1/2} =\pm m$  is used. In this {case, the} operators,  $\hat M_{\pm}=\hat M_x\pm i \hat M_y$, 
connecting 
wave functions {and} %with 
$m\to m\pm 1$,  can be defined {and,} acting by $\hat M_{\pm}$, results in a set of eigenfunctions with the eigenvalues \cite{JKT1}, %: 
\begingroup
\makeatletter\def\f@size{8}\check@mathfonts
\def\maketag@@@#1{\hbox{\m@th\normalsize\normalfont#1}}%
\begin{eqnarray}
\{ m_1\}\downarrow_{(m^2)^{1/2} =\pm m} &=& (L-2 k_1) = \{ L; L-2;  \ldots ; L-2[L/2]; L-2[L/2]-2; \ldots ;  -\infty \} ,  \nonumber\\
\{ m_1\} \uparrow_{(m^2)^{1/2} =\pm m} &=& (-L+2 k_2) =  \{-L;  -L+2;  \ldots ; -L+2[L/2];  -L+2[L /2] +2 ; \ldots ; \infty  \} , \nonumber\\
\{ m_2\} \downarrow_{(m^2)^{1/2} =\pm m} &=& (L-1-2 k_3) =  \left\{L-1;  L-3;  \ldots ; \right. \nonumber\\
 	       &&  \left. L-1-2[(L-1)/2]; L-1-2[(L-1) /2]-2; \ldots ; -\infty  \right\} , \nonumber\\
\{ m_2\}\uparrow_{(m^2)^{1/2} =\pm m}  &=& (-L+1+2 k_4) =  \left\{ -L+1; -L+3;  \ldots ; \right. \nonumber\\ 
               && \left.-L+1+2[(L-1)/2]; -L+1+2[(L-1) /2]+2; \ldots ; \infty \right\} , 
\label{eq34}
\end{eqnarray} 
\endgroup  
where  $k_1, k_2,k_3,k_4$ are positive integers and the elements of the sets $\{ m_j\}$ can be any real number, not necessarily integer or half-integer.  
{For %the 
$m$s}  %MDPI: looks plural. Plese confirm.
from {Equation} 
(\ref{eq34}),  the corresponding hypergeometric {functions,} ${}_2F_1$, are regular, some 
$f(m|\theta)\sim (\sin^2\theta) ^ 
{(m^2)^{1/2}/2}=(\sin\theta)^{\pm m}$ 
 %become 
{are} 
singular {and,} therefore,  some $\Psi_{M(1,2)}(L,m|\theta)$ also %become
 {are} %MDPI: please confirm teh replacements as the meaning retained. 
singular{; %(
for explicit calculations 
and technicalities,} %technical details, 
see \cite{JKT1}. %).  
For the case {of} %MDPI: fits better, please check all text.
$(m^2)^{1/2}=\pm m$, %we obtain, 
{one obtains,} 
{similar to  %to the result %MDPI: extra wording, clear enough.
 %for 
the spectrum, resulting} %following %MDPI: rewritten, please confirm the meaning retained ("following to" vs. "resulting from") 
from Equations~(\ref{eq31}) and  (\ref{eq310}), 
that the spectrum is discrete with the only condition that $L-|m|$ is necessarily {integer,} while $L$ and $m$ 
 %separately could 
{can}
be integer as  well as 
non-integer {each.} %MDPI: rewritten, please confirm the meaning retained.
 %When 
 {If} 
$L$ is {integer,} the  sequences (\ref{eq34}) do not extend to $\pm \infty$ but truncate at $\pm L$ and reproduce the set of eigenvalues (\ref{eq1.20}), 
obtained using  {prescription,} $(m^2)^{1/2}=|m|$.   

Let us note that in the group theoretical {framework, the eigenfunctions,} corresponding to the eigenvalue spectrum  (\ref{eq34}), form an irreducible 
representation of  $SO(3)$, the three dimensional rotation group{; %(
 see,} e.g., \cite{group}. %). 
 As mentioned {just} above, %MDPI: otherwise to precise in which Section. please confirm.
when $L$ and $m$ are integer, infinite sequences (\ref{eq34}) truncate into a finite set of eigenvalues, 
$-L\leq m \leq L$, and the corresponding eigenfunctions are regular and form a finite set. In terms of the representation theory, these are the finite 
dimensional irreducible representations of the $SO(3)$ group. When  $L$ and $m$ are non integer (half-integers included), the  sequences (\ref{eq34}) do not 
truncate and remain {infinite. 
 %and 
 Then,} %MDPI: two sentences instead of one. Please confirm teh emaning retained.
the corresponding infinite set of eigenfunctions is formed 
 %from 
 {of} %MDPI: fits better.
both singular and regular functions. In terms of the representation theory, these are the infinite dimensional irreducible representations of the $SO(3)$ group. Using prescription $(m^2)^{1/2} =\pm m$ results  in an infinite set of eigenfunctions
 %, 
{(containing both singular and regular functions),}  %MDPI: put in the brakets as teh secondary comment, not to break sense, please confirm.
corresponding to an infinite dimensional representation of the rotation group. 
 %When using 
{Using} %MDPI: fits better. Rewritten, please confirm.
$(m^2)^{1/2} 
=|m|$, the finite set of eigenvalues $-L\leq m \leq L$, symmetric with respect to $m\rightarrow \,-m$, is filtered out from the infinite set (\ref{eq34})
{because of $|m\geq 0$.} 
The set of corresponding eigenfunctions is comprised of regular functions only. In other words, the set of regular eigenfunctions is being filtered out from a general infinite set of eigenfunctions exactly the same 
 %was
 {way, %MDPI: looks wrong wording. please confirm the replacement and changes. 
similar to the
 %as in 
case 
 %when
 of integer} 
$L$ and $m$. % are integer.

So, depending on a prescription for $(m^2)^{1/2}$, eigenfunctions of the operator of angular momentum square could be regular or singular and the eigenvalues could be given either \mbox{by 
{Equations} 
(\ref{eq1.20})} or (\ref{eq34}). When the {prescription,}  $(m^2)^{1/2} =|m|$, is used, {all} %the 
eigenfunctions   are regular and the eigenvalue spectrum 
is 
given by Equation~(\ref{eq1.20}). When the prescription is  $(m^2)^{1/2} =\pm m$, some eigenfunctions  are regular, {while}  %and 
some {are} singular and the eigenvalue spectrum is given by Equation~(\ref{eq34}). 

 %At the end, 
{Finally,} 
let us discuss what could cause the statement that the eigenvalue problem for the $\hat M^2$  admits normalizable solutions only when $L$ is integer \cite{Bohm,Schiff,Fock,LL}. The eigenfunctions and the spectrum, e.g., the set of regular functions and eigenvalues (\ref{eq1.20}), are obtained by requiring normalizability of a solution that is presented in terms of a specific pair of linearly independent functions, $\Psi_{M(1)}$ and $\Psi_{M(2)}$. Quite a different picture arises when the  normalization condition is applied to another pair of linearly independent functions, e.g., 
{to} %MDPI: missing word.
the Legendre functions, $P^m_L(\theta)$  and $Q^m_L(\theta)$, which were, from the  early days of  quantum mechanics, considered as eigenfunctions of the operator  of the angular momentum squared  \cite{Bohm,Schiff,Fock,LL}.

%Of course 
{Certainly,} %MDPI: fits better. 
both pairs, $\Psi_{M(1)}(L,m|\theta)$, $\Psi_{M(2)}(L,m|\theta)$ and $P^m_L(\theta)$, $Q^m_L(\theta)$, are solutions of the eigenvalue 
 %Equation 
 {equation} %MDPI: here no cap[ital "E". 
(\ref{M2}). Legendre functions can be  written as linear combinations of hypergeometric functions, $\Psi_{M(1,2)}(L,m|\theta)$ \cite{ww}:
\begin{eqnarray}
P^m_L(\theta)& = &C_{11}  \Psi_{M(1)}(L,m|\theta) + C_{12} \Psi_{M(2)}(L,m|\theta);\nonumber\\
Q^m_L(\theta) &=& C_{21}  \Psi_{M(1)}(L,m|\theta) + C_{22} \Psi_{M(2)}(L,m|\theta).
\label{eq36}
\end{eqnarray}

{Using} just a polynomialization, it is  not {sufficient} %enough %MDPI: fits better. 
to normalize both $ P^m_L(\theta) $ and $Q^m_L(\theta) $ simultaneously.   The {reason} %for this  %MDPI: unneeded wording.
is that 
the sets $\{ m_{(1)}\}$ and $\{ m_{(2)}\}$, generated by the two conditions of polynomialization (\ref{eq31}) and (\ref{eq310}), have no 
%{a} %MDPI: here means "no one". please confirm teh sense.
common element, {and, thus,} it is impossible to satisfy  these  conditions simultaneously. %When using   
{As soon as the} %MDPI: fits better.
polynomialization conditions {are applied}, either 
$\Psi_{M(1)}(L;m|\theta)$ or   $\Psi_{M(2)}(L;m|\theta)$ {is} 
 %always %MDPI: no needed, clear enough.
 singular \cite{JKT1}.  Therefore, in order to carry out {the} 
normalizability of $ P^m_L(\theta) $ and $Q^m_L(\theta) $, only polynomialization would not 
suffice and  an additional requirement {to filter} %, filtering 
out singular parts %from
 {of}  
 $ P^m_L(\theta) $ and $Q^m_L(\theta) %, %MDPI: rewritten; fits better.
 $ is necessary.  This can be achieved by choosing {the 
coefficients,} $C_{ij}$ in {Equation} (\ref{eq36}). Namely, {as soon as} %when 
 the 
polynomialization condition (\ref{eq31}) is satisfied, {what  leads} 
 %leading 
to %a 
 regular $\Psi_{M(1)}(L;m|\theta)$ and singular $\Psi_{M(2)}(L;m|\theta)$, {the 
coefficients,} $C_{12}$ and $C_{22}$, 
of $\Psi_{M(2)}(L;m|\theta)$ %, 
must vanish.
Similarly, %when
 {as soon as}  
the polynomialization condition (\ref{eq310}) is satisfied %, 
 {what leads} 
 %leading 
to  %a 
regular $\Psi_{M(2)}(L;m|\theta)$ and singular $\Psi_{M(1)}(L;m|\theta)$, 
{the} 
coefficients, 
$C_{11}$ and $C_{21}$, 
of $\Psi_{M(1)}(L;m|\theta)$ %, 
must vanish.

Coefficients $C_{ij}$ %were
 {are}  
calculated in \cite{JKT1}:
\begingroup
\makeatletter\def\f@size{9}\check@mathfonts
\def\maketag@@@#1{\hbox{\m@th\normalsize\normalfont#1}}%
\begin{flalign}\nonumber
C_{11}^{-1}\sim\Gamma\left({1\over 2}-{L\over 2}+{|m|\over 2}\right) \,\Gamma\left(1+{L\over 2}+{|m|\over 2}\right),\quad C_{12}^{-1}\sim\Gamma\left({1\over 2}+{L\over 2}+{|m|\over 2}\right) \Gamma\left(-{L\over 2}+{|m|\over 2}\right),\end{flalign}
\begin{equation}
C_{21}\sim{\Gamma\left({1\over 2}+{L\over 2}-{|m|\over 2}\right) \over\Gamma\left(1+{L\over 2}+{|m|\over 2}\right)},\quad 
C_{22}\sim{\Gamma\left(1+{L\over 2}-{|m|\over 2}\right) \over \Gamma\left({1\over 2}+{L\over 2}+{|m|\over 2}\right)}.
\label{Tesla}
\end{equation}
\endgroup  
 {Here} $\Gamma$ is the
Euler Gamma function. %MDPI: the variables to be defined as met first time. 

{It} {is straightforward to show  {that,} after applying {the} polynomialization \mbox{conditions (\ref{eq31})}} \mbox{and 
(\ref{eq310})} {and  using the property of the 
 %Euler 
$\Gamma$ function {such}  
that \mbox{$1/\Gamma(\mathrm{nonpositive}\;\;\mathrm{integer})=0$ \cite{ww,Abramowitz}}}, 
 %we obtain 
{one obtains} 
that $C_{11}=0$ and $C_{12}=0$ can be realized only {for %the 
integer} %s 
 $L$ and $m$ and that $C_{21}=0$ and $C_{22}=0$ can never be satisfied \cite{JKT1}.  Therefore, $Q^m_L(\theta)$ has to be excluded, and only 
$P^m_L(\theta)$  remains as the quantum mechanical eigenfunction. Consequently, for the pair $P^m_L(\theta)$ and $Q^m_L(\theta)$,  normalizability is achieved only when $L$ and $m$ are integer.  This is not a general result;  one explicit example of  the eigenfunction, normalizable for integer as well as for non-integer eigenvalues,  is presented by the pair $\Psi_{M(1)}(L,m|\theta)$ and $\Psi_{M(2)}(L,m|\theta)$, given in  
{Equation} 
(\ref{eq30}) and leading %, 
to spectra (\ref{eq1.20}) or (\ref{eq34}),
{depending on the prescription for the power function.} %MDPI: fits better placed here. 
  {Hence,}  the statement that from theoretical quantum mechanics it follows that the eigenvalue spectrum of $\hat 
M^2$  is comprised of only integers is not necessarily correct in %the 
{sense} that it corresponds to a special case when $P^m_L(\theta)$ and $Q^m_L(\theta)$  are chosen as a pair of linearly independent solutions of the 
eigenvalue 
 %Equation 
{equation}
(\ref{M2}). 

\section{4. Conclusions}
\label{s4}

 %We studied 
{In this paper,}
the eigenvalue problem for the operator of the angular momentum {is studied} 
in the framework of {nonrelativistic} quantum mechanics. The general result for the spectrum is that it is discrete, {namely,}  $|m|=L-k$ with $k$ 
{been} integer, 
$k=\{0,1,\cdots , [L] \}$, where $[L]$ is the integer part of $L$. $L$ and $m$ can be integer as well as non-integer and this does not contradict 
 %to 
{any} physical principle. 

The above is in stark contrast with the %well-
 {known} 
statement that from theoretical quantum mechanics it follows that $m$ and $L$ {can 
only be}  
integer or half-integer \cite{Bohm,Schiff,Fock,LL}.  {An %E
 explanation} of {the} %is 
contradiction is that 
 %our 
{the}
result, %is 
obtained {here, does not impose a} 
 %without imposing 
non-physical  requirements of either periodicity of the wave function  or postulating that,  moving  with the step size 1 and starting from a state with $m=-L$, 
 %we 
{one} 
should arrive at the state with $m=+L$ and vice versa.  The discreteness {condition,} $|m|=L-k$, does not require that moving with the step size 1 from 
$\Psi_M(L,-L|\theta)$ {one} should {end with} %MDPI: rewritten to fit better. please confirm the meaning. 
 %lead  to 
$\Psi_M(L,+L|\theta)$.

Using  the Legendre functions, $P^m_L(\theta)$ and $Q^m_L(\theta)$, as a pair of linearly independent solutions for the eigenvalue equation, $\hat M^2 
\Psi=L(L+1)\Psi$,  is a specific choice that does not encompass  the most general case. When  $P^m_L(\theta)$ and $Q^m_L(\theta)$  are used, {the}  
normalizability requirement filters out non-integer 
{$L$ and $m$,} 
%$L,\,m$, 
but 
 %it may exist solution of 
the eigenvalue equation {solution} 
that is normalizable {may exist} 
for any real 
eigenvalues, integer and non-integer. %MDPI: rewritten. please check teh meaning retained. 
 %We have presented a
{Another} solution, presented here, Equation~(\ref{eq30}), %which %MDPI: fits better.
satisfies the necessary physical requirement of {the} normalizability for integer 
and non-integer 
 %$L,\,m$.
{$L$ and $m$.}

Imposing the condition of the single valuedness on the eigenfunction of the third component of the angular momentum $(x+i y)^m$ does not necessarily lead 
to the violation of the rotational invariance. Indeed, 
 %we have found 
{it is found} 
such a representation of the power function, Equation~(\ref{eq10}), which for any $m$ is a single valued function of 
 {($x,\,y$)} %MDPI: presented a apair, otherwise to write "x and y". 
and is invariant under the {$2\pi k$ rotation.} %  at $2\pi k$. 
Two results indicate that $\Psi(x,y|m)$ from  Equation~(\ref{eq10}) and  $(x+iy)^m$ belong to the same class of functions. 
{First,} %ly, 
according to Equation~(\ref{eq11}), solution $\Psi(x,y|m)$, {coincide,} %MDPI: fits beter here. 
up to the normalization factor, %coincides 
with $(x+iy)^m$ when $m$ is integer. 
 {Second,} %ly, %MDPI: fits better.
 for the half-integer values of $m=N/2$, 
 %we obtained
 {one obtaines}
 $(\Psi(x,y|N/2))^2\sim \Psi(x,y|N)$ (Equation~(\ref{eq18})), 
relation mirroring the one valid for the function (\ref{eq2}), $((x+iy)^{N/2})^2=(x+iy)^N$. For the rational $m=p/n$, 
 %we have 
{it is}
not shown that $(\Psi(x,y|p/n))^n\sim \Psi(x,y|p)$ but  particular cases of integer and half-integer $m$ are 
 %signs 
 {indications} %MDPI: fits better.
that $(\Psi(x,y|\alpha))^\beta\sim \Psi(x,y|\alpha\beta)$ may 
 %very %MDPI: No "very", "absolutely", etc. are recomemnded for scientific papers.
{well enough} be true for any 
 %$\alpha,\;\beta$.  
 {$\alpha$ and $\beta$.}  
Another important {property,} indicating that $(x+iy)^m$ and $\Psi(x,y|m)$ belong to the same {class,}
is that 
for 
any $m$,  %they 
both 
{the functions} 
satisfy {the} relation, $|\Psi/C|^2=1${; %, 
 see} \mbox{Equation~(\ref{eq25})}. 
 %We conclude 
{To summarize,} %MDPI: or "To conclude" 
 %that 
the wave function (\ref{eq10}), solution of the eigenvalue equation $\hat M_z\Psi(x,y|m)=m\, \Psi(x,y|m)$, 
{represents a} %presents one 
possible  prescription for the power {function,} $(x+i y)^m$.  

In the general case, when the only condition imposed on a wave function is the physical requirement of {the} normalizability, i.e., when 
the  periodicity requirement for a wave function is lifted or when the different pair of linearly independent functions is chosen, 
there is no constraint on $L$ and $m$ to be  %only 
integer {only.} From %the physical  standpoint, 
 {the physics point of view,} %MDPI: looks this is what is meant?
 the only  self-consistent approach 
is 
to drop {all} %the 
non-physical conditions and consider the problem in the presence of only the physical requirements. 
This is what {is} %was %MDPI fits better. 
done in %our  work
 {this paper}
{and,} as a result, %we obtained 
a new quantum-mechanical solution of the eigenvalue problem for the angular momentum operator {is obtained}.

To conclude, the main result of this paper is that from the framework of theoretical quantum mechanics it does not follow that the eigenvalues of the 
angular momentum operator should {only} be %only 
integer. 

 %Of course, 
{Surely,} %MDPI: fits better.
the spectrum of the angular momentum cannot be defined from theoretical quantum mechanics alone but has to be established by comparing 
theoretical calculations with experiments. %This
 {However, this}  
is not a goal of the %present
 {current}  
paper which seeks to analyze the eigenvalue problem for the angular momentum 
operator from the purely theoretical {viewpoint.} %standpoint. %MDPI: fits better, please confirm.

\acknowledgments{We are indebted with C.~M.~Bender,  E.~E.~Boos, O.~Daalhuis, J.~T.~Gegelia and V.~A.~Petrov for illuminating discussions.}

%\conflictsofinterest{The authors declare no conflict of interest.}%MDPI:  Declare conflicts of interest or state ``The authors declare no conflict of interest.'' Authors must identify and declare any personal circumstances or interest that may be perceived as inappropriately influencing the representation or interpretation of reported research results. Any role of the funders in the design of the study; in the collection, analyses or interpretation of data; in the writing of the manuscript, or in the decision to publish the results must be declared in this section. If there is no role, please state ``The funders had no role in the design of the study; in the collection, analyses, or interpretation of data; in the writing of the manuscript, or in the decision to publish the~results''.

%\begin{adjustwidth}{-\extralength}{0cm}
%\reftitle{References}

% Please provide either the correct journal abbreviation (e.g., according to the “List of Title Word Abbreviations” http://www.issn.org/services/online-services/access-to-the-ltwa/) or the full name of the journal.
% Citations and References in Supplementary files are permitted provided that they also appear in the reference list here. 

%=====================================
% References, variant A: external bibliography
%=====================================
%\bibliography{your_external_BibTeX_file}

%=====================================
% References, variant B: internal bibliography
%=====================================
\begin{thebibliography}{999}
\bibitem{Bohm}
Bohm, D. {\it Quantum Theory}; Dover {Publications, Inc.: New York, NY, USA,} %MDPI: Newly added information, please confirm. The same for below.
1989.
\bibitem{Schiff}
Schiff, L.{I.}  {\it Quantum Mechanics}; McGraw-Hill {Book Company: New York, NY, USA}, 1968.
\bibitem{Fock} 
Fock, V.A.  {\it Fundamentals of Quantum Mechanics}; Mir Publishers: {Moscow, USSR}, 1970.
\bibitem{LL}
 Landau, L.D.; Lifshits, E.M. {\it Quantum Mechanics: Non-Relativistic Theory}; Butterworth-Heinemann: {Oxford, UK}, 1991.%\hyperref[Landau{''https://en.wikibooks.org/wiki/LaTeX/''}
  
\bibitem{M}
Majorana, E. {\it Teoria relativistica di particelle con momento intrinseco arbitrario}; {Nuovo Cimento} {\bf 1932}, {9}, {335--344}.\\{https://doi.org/10.1007/BF02959557}

\bibitem{Regge}
de Alfaro, V.; and Regge, T. {\it Potential Scattering}; {North-Holland Publishing Company: Amsterdam, North Holland,} 
1965.

\bibitem{GG}
G$\ddot{o}$tte, J.~B.; Franke-Arnold, S.; Zambrini, R.; Barnett, S.M. {\it Quantum formulation of fractional orbital angular momentum;} 
{J. Mod. Opt.} {\bf {2007}},
{54}, 1723{--1728}.
\\{https://doi.org/10.1080/09500340601156827}

\bibitem{Land}
Land, M. {\it Harmonic oscillator states with integer and non-integer orbital angular momentum;} 
{J.~Phys. Conf. Ser.} {\bf {2011}},  {330}, 012014.
\\{https://doi.org/10.1088/1742-6596/330/1/012014}

\bibitem{JKT1}
Japaridze, G.S.; Khelashvili, A.A.; Turashvili, K.S. {\it Critical comments on quantization of the angular momentum: I. Analysis based on the physical requirement on eigenfunctions and on the commutation relations;} %MDPI: Newly added information, please confirm.
arXiv:1912.08042.
\\{https://doi.org/10.48550/arXiv.1912.08042} %MDPI: very recently arXiv got DOI.

\bibitem{JKT2}
Japaridze, G.S.; Khelashvili, A.A.; Turashvili, K.S. {\it Critical comments on the quantization of the angular momentum: II. Analysis based on the requirement that the eigenfunction of the third component of the operator of the angular momentum must be a single valued periodic function.}; arXiv:2004.10673. \\{https://doi.org/10.48550/arXiv.2004.10673}


\bibitem{ww}
Whittaker, E.T.; Watson, G.N.  {\it A Course of Modern Analysis}; Cambridge University Press: {Cambridge, UK}, 1996.\\{https://doi.org/10.1017/CBO9780511608759}

\bibitem{Pauli}
{Pauli,} W. {\it\"Uber ein Kriterium f\"ur Ein- oder Zweiwertigkeit der Eigenfunktionen in der Wellenmechanik.} 
{ Helv. Phys. Acta}  {\bf 1939}, { 12}, 147{--168}, 
\\{available online:} 
\url{https://www.e-periodica.ch/digbib/view?pid=hpa-001%3A1939%3A12#157} %(accessed on 15 May 2022).
;\\ reprinted in {\it Collected Scientific Scientific Papers by W.Pauli, in two Volumes", Ed. By 
R.~Kronig and V.F.~Weisskopf}, Interscience Publishers, John Wiley and Sons, Inc. New York, NY, USA, 1964.
 %MDPI: A Reference can only have one journal. We removed second ref Moreover it does not have pages and Volume.
 % If you have the pages and volume, please add this as second ref. Please also give the title. 



 \bibitem{merz}
Merzbacher, E.  {\it Single valuedness of wave functions;}
{Am. J. Phys.} {\bf {1962}}, %MDPI: Please provide article title.
{30}, 237{--247}.
 \\{https://doi.org/10.1119/1.1941984}

\bibitem{Abramowitz}
Abramowitz, M.; Stegun, I.{A., Eds.}  
{\it Handbook of Mathematical Functions}; Dover {Publications, Inc.:} New York, {NY, USA}, 1970.

\bibitem{Br}
Prudnikov, A.P.; Brychkov, Y.A.; Marichev, O.I.  {\it Integrals and Series. Volume 3: More Special Functions}; {Gordon and Breach Science Publishers: 
London, UK} 
1992.

\bibitem{NIST}
 Olver, F.W.J.; {Lozier, D.W.; Boisvert, R.F.; Clark, C.W., Eds.  } %MDPI: please confirm.
{NIST Digital Library of Mathematical Functions}. Comments to Formula 15.4.12.  Available online: 
\\{https://dlmf.nist.gov}  {(accessed on 15 May 2022).} %MDPI: accessed date added.


\bibitem{group}
Vilenkin, N.{Ja.} {\it Special Functions and the Theory of Group Representations}; 
 %Translations of Mathematical Monographs; 
{American Mathematical Society: Providence, RI, USA}
1968.   Volume 22. \\{https://doi.org/10.1090/mmono/022} 

%A.~A.~Kirillov, {\it Elements Of The Theory Of Representations}, Springer, (1976).
\end{thebibliography}

% If authors have biography, please use the format below
%\section*{Short Biography of Authors}
%\bio
%{\raisebox{-0.35cm}{\includegraphics[width=3.5cm,height=5.3cm,clip,keepaspectratio]{Definitions/author1.pdf}}}
%{\textbf{Firstname Lastname} Biography of first author}
%
%\bio
%{\raisebox{-0.35cm}{\includegraphics[width=3.5cm,height=5.3cm,clip,keepaspectratio]{Definitions/author2.jpg}}}
%{\textbf{Firstname Lastname} Biography of second author}

% For the MDPI journals use author-date citation, please follow the formatting guidelines on http://www.mdpi.com/authors/references
% To cite two works by the same author: \citeauthor{ref-journal-1a} (\citeyear{ref-journal-1a}, \citeyear{ref-journal-1b}). This produces: Whittaker (1967, 1975)
% To cite two works by the same author with specific pages: \citeauthor{ref-journal-3a} (\citeyear{ref-journal-3a}, p. 328; \citeyear{ref-journal-3b}, p.475). This produces: Wong (1999, p. 328; 2000, p. 475)

%%%%%%%%%%%%%%%%%%%%%%%%%%%%%%%%%%%%%%%%%%
%% for journal Sci
%\reviewreports{\\
%Reviewer 1 comments and authors’ response\\
%Reviewer 2 comments and authors’ response\\
%Reviewer 3 comments and authors’ response
%}
%%%%%%%%%%%%%%%%%%%%%%%%%%%%%%%%%%%%%%%%%%
%\end{adjustwidth}
%\end{adjustwidth}
\end{document}